\documentstyle[12pt,epsfig]{article}
\let\section=\subsection     \let\subsection=\subsubsection                

\def\tr{{\rm tr} \,}

\begin{document}
\begin{center}
   {\large \bf Effective chiral theory of kaon-nucleon scattering}\\[2mm]
      M.F.M.~LUTZ and E.E.~KOLOMEITSEV\\[5mm]
   {\small \it Gesellschaft f\"ur Schwerionenforschung (GSI) \\
   Planckstr. 1, D-64291 Darmstadt, Germany  }
\end{center}

\begin{abstract}\noindent
We apply the relativistic chiral $SU(3)$ Lagrangian density to kaon-nucleon 
scattering imposing constraints from the pion-nucleon s- and p-wave 
threshold parameters at chiral order $Q^2$. The s and u
channel decouplet baryon exchange is included explicitly and is found to play a 
crucial role in understanding the empirical s- and p-wave nuclear kaon dynamics 
quantitatively. 
\end{abstract}

\section{Introduction}

A good understanding of the $\bar K$-nucleon interaction is required for the 
description of $K^-$-atoms \cite{Gal} and the subthreshold production of kaons in heavy ion 
reactions \cite{Senger}. The ultimate goal is to relate the in-medium spectral function of 
kaons with the anticipated chiral symmetry restoration at high baryon density. 

In early approaches  the $\bar K$-nucleon scattering amplitudes are described in terms of 
a coupled channel $K$-matrix with the set of parameters adjusted to available 
elastic and inelastic cross sections at small laboratory momenta $p_{\rm lab}\leq 200 $ MeV.
Due to the insufficient quality of the low energy elastic and inelastic $K^-p$ data the 
resulting scattering amplitudes remained ambiguous. The sign of the $K^-$-proton scattering 
length was only recently convincingly determined by a kaonic-hydrogen-atom measurement 
\cite{Iwasaki}. The sign of the $K^-$-neutron scattering length remains highly 
model dependent \cite{A.D.Martin,martsakit}. This reflects the fact that there are no 
$K^-$ neutron (deuteron) scattering data available at low energies and therefore the isospin 
one scattering amplitude is constrained only indirectly for example by 
the $\Lambda \pi^0$ production data. As a consequence also the subthreshold $\bar K N$ scattering amplitudes, which determine 
the $\bar K$-spectral function in nuclear matter to leading orders in the density expansion, 
are poorly known. In the region of the $\Lambda^*(1405)$ resonance the isospin zero 
amplitudes of various analyses may differ by a factor of two \cite{Kaiser,Ramos}.
Here a profound theory is asked for. In particular it is desirable to utilize the chiral 
symmetry constraints of QCD. First attempts in this directions can be found in 
\cite{Kaiser,Ramos}. The reliability of the extrapolated subthreshold scattering amplitudes 
can further be substantially improved by including both s- {\it and} p-waves in the analysis 
of the empirical cross sections since for $p_{\rm lab}>200$ MeV the available data are much 
more precise than at $p_{\rm lab}<200$ MeV where one expects s-wave dominance. 

In this work we apply the relativistic chiral $SU(3)$ Lagrangian including 
an explicit baryon decouplet field. As to our knowledge 
this is the first application of the chiral $SU(3)$ Lagrangian to the s- and p-wave 
nuclear kaon dynamics. It is argued that due to 
the rather large kaon mass the kaon-nucleon dynamics turns non-perturbative in contrast 
with the pion-nucleon system where chiral perturbation theory can be applied successfully 
\cite{Bernard}. This implies that a partial summation scheme is required in the strangeness 
sector \cite{njl-lutz,Kaiser,Ramos}. We solve the Bethe-Salpeter equation for the scattering 
amplitude with the interaction kernel truncated at chiral order $Q^2$. This leads to s- 
and p-wave contributions in the scattering amplitude. 
As a crucial technical ingredient we propose to 
apply a subtraction scheme rather than a cutoff scheme as employed in \cite{Kaiser,Ramos}. 
The subtraction point is identified with the hyperon mass as to protect the s-channel 
hyperon exchange term contributing in the $P_{11}$ and $P_{31}$ channels. The 
renormalization scheme is an important input of our chiral $SU(3)$-dynamics. In particular 
we avoid first, an uncontrolled breaking of the $SU(3)$-symmetry induced by channel dependent 
cutoff parameters as in \cite{Kaiser} and second, a strong sensitivity of the 
$\Lambda^*(1405)$ resonance structure on the cutoff parameter implicit in \cite{Kaiser,Ramos}. 

We successfully adjust the set of parameters to describe the existing $K^-\,p$ elastic and 
inelastic cross section data including angular distributions to good accuracy. 
Moreover the measured spectral form of the $\Lambda^*(1405)$ and $\Sigma^*(1385)$ resonances 
are reproduced. As a result of our analysis we find a weakly repulsive s-wave 
$K^-$ neutron interaction and a subthreshold $K^-$ nucleon forward scattering 
amplitude with sizable contributions from p-waves. 

\section{Relativistic chiral $SU(3)$ interaction terms}

We recall the relevant interaction terms of the relativistic chiral
$SU(3)$ Lagrangian density. The systematic construction principle can be found in 
\cite{Krause}. A systematic regrouping of interaction terms accompanied by an appropriate 
subtraction scheme leads to manifest chiral power counting rules \cite{nn-lutz}. At chiral 
order $Q^2$ we collect the relevant interaction terms:
\begin{eqnarray}
{\mathcal L}&=&   
\tr \bar B \left(i\,\partial \!\!\!/-m_B\right) \, B 
+\frac{1}{4}\,\tr (\partial^\mu \,\Phi )\,
(\partial_\mu \,\Phi )
+\frac{i}{8\,f_\pi^2}\,\tr\bar B\,\gamma^\mu \Big[\Big[ \Phi , 
(\partial_\mu \,\Phi) \Big]_-,B \Big]_-
\nonumber\\
&+&\frac{F}{2\,f_\pi} \,\tr  \bar B
\,\gamma_5\,\gamma^\mu \,\Big[\left(\partial_\mu\,\Phi\right),B\Big]_-
+\frac{D}{2\,f_\pi} \,\tr  \bar B
\,\gamma_5\,\gamma^\mu \,\Big[\left(\partial_\mu\,\Phi\right),B\Big]_+
\nonumber\\
&+&\tr \bar \Delta_\mu \cdot \Big(
\left( i\,\partial \!\!\!/ -m_\Delta \right)g^{\mu \nu }
-i\,\left( \gamma^\mu \partial^\nu + \gamma^\nu \partial^\mu\right)
+i\,\gamma^\mu\,\partial \!\!\!/\,\gamma^\nu
+\,m_\Delta \,\gamma^\mu\,\gamma^\nu
\Big) \, \Delta_\nu
\nonumber\\
&+&\frac{C}{2\,f_\pi}\,
\tr \left\{ 
\Big( \bar \Delta_\mu \cdot 
(\partial_\nu \,\Phi ) \Big)
\Big( g^{\mu \nu}-{\textstyle{1\over2}}\,Z\, \gamma^\mu\,\gamma^\nu \Big) \,
B +\mathrm{h.c.}
\right\}
\nonumber\\
&+&\frac12\,g^{(S)}_0\,\tr\bar B\,B
\,\tr (\partial_\mu\Phi) \, (\partial^\mu\Phi)
+\frac12\,g^{(S)}_1\,\tr \bar B \,(\partial_\mu\Phi)
\,\tr(\partial^\mu\Phi) \, B 
\nonumber\\
&+&\frac14\,g^{(S)}_F\, \tr\bar B \Big[
\Big[(\partial_\mu\Phi),(\partial^\mu\Phi)
\Big]_+ ,B\Big]_- 
+ \frac14\,g^{(S)}_D\,\tr \bar B \Big[
\Big[(\partial_\mu\Phi),(\partial^\mu\Phi)
\Big]_+, B \Big]_+ 
\nonumber\\
&+&\frac{1}{4}\,g^{(V)}_0\,
\Big(\tr \bar B \,i\,\gamma^\mu\,( \partial^\nu B) \,
\tr(\partial_\nu\Phi) \, ( \partial_\mu\Phi) 
+\mathrm{h.c.}\Big)
\nonumber\\
&+&\frac{1}{8}\,g^{(V)}_1\,
\Big(\tr \bar B \,( \partial_\mu\Phi)
\,i\,\gamma^\mu\,
\tr(\partial_\nu\Phi) \, ( \partial^\nu B) 
+\mathrm{h.c.}\Big)
\nonumber\\
&+&\frac{1}{8}\,g^{(V)}_1\,
\Big(
\tr \bar B \, ( \partial_\nu\Phi)
\,i\,\gamma^\mu\,
\tr ( \partial^\nu \Phi) \, ( \partial_\mu B)
+\mathrm{h.c.}\Big)
\nonumber\\
&+&\frac{1}{8}\,g_F^{(V)}\,\Big(
\tr   \bar B \,i\,\gamma^\mu\,\Big[
\Big[(\partial_\mu \Phi) , (\partial_\nu\Phi)\Big]_+,
( \partial^\nu B) \Big]_-
+\mathrm{h.c.} \Big)
\nonumber\\
&+& \frac{1}{8}\,g^{(V)}_D\,\Big(\tr 
\bar B \,i\,\gamma^\mu\,\Big[
\Big[( \partial_\mu\Phi) , (\partial_\nu\Phi)\Big]_+, 
( \partial^\nu B) \Big]_+
+\mathrm{h.c.} \Big)\,,
\nonumber\\
&+&
\frac{1}{2}\,g^{(T)}_1\,\tr\bar B \,( \partial_\mu\Phi)
\,i\,\sigma^{\mu \nu}\,\tr( \partial_\nu \Phi) \, B 
\nonumber\\
&+& \frac{1}{4}\,g^{(T)}_D\,
\tr \bar B \,i\,\sigma^{\mu \nu}\,\Big[
\Big[(\partial_\mu \Phi) ,( \partial_\nu \Phi)  \Big]_-, B \Big]_+
\nonumber\\
&+&\frac{1}{4}\,g^{(T)}_F\, 
\tr \bar B \,i\,\sigma^{\mu \nu}\,\Big[
\Big[( \partial_\mu \Phi) ,( \partial_\nu\Phi) \Big]_- ,B\Big]_-
\label{lag-Q}
\end{eqnarray}
where we introduce the meson octet field $\Phi=\sum \Phi_i\,\lambda_i$,  
the baryon octet field $B= \sum B_i\,\lambda_i/\sqrt{2}$ and the
completely symmetric  baryon decouplet field $\Delta $. The parameter
$f_\pi \simeq 93$ MeV is determined from the weak decay width of charged pions.
Similarly the parameters  $F\simeq 0.45$ and $D\simeq 0.80$ are constrained by the 
weak decay widths of the baryon octet states \cite{Okun:1982ap} and $C\simeq 1.5-1.7$ is 
constrained by the hadronic decay width of the baryon decouplet states. We point out that 
the parameter $Z$ in (\ref{lag-Q}) can be reliably determined only in an $SU(3)$ analysis 
of meson-baryon scattering since in the $SU(2)$ sector its effect can be absorbed into 
the local 4-point interaction terms. The remaining 11 free parameters of the local 4-point 
interaction terms describe s-wave range and p-wave dynamics. If the heavy baryon expansion 
is applied 
to (\ref{lag-Q}) the local 4-point interaction terms can be mapped onto corresponding terms
of the heavy baryon formalism presented for example in \cite{CH-Lee}. However, we prefer 
the manifestly covariant form (\ref{lag-Q}). The interaction terms of (\ref{lag-Q}) predict 
well defined kinematical structures which we find crucial for a quantitative description of 
the nuclear kaon dynamics.

Further interaction terms are induced by the explicit  chiral symmetry breaking  
of QCD \cite{Manohar,Kaiser}:
\begin{eqnarray}
{\mathcal L }_{\chi-SB}&=& b_D\,\tr \bar B \,\Big[ \chi , B \Big]_+ +b_F\,\tr
\bar B \,\Big[ \chi , B \Big]_-
+b_0 \,\tr\bar{B}\, B \,\tr\chi
\nonumber\\
&+&d_D\,\tr \Big(\bar \Delta_\mu\cdot \Delta^\mu  \Big)  \,\chi
+d_0 \,\tr \left(\bar \Delta_\mu\cdot  \Delta^\mu\right) \,\tr\chi 
\,,
\nonumber\\
\chi &=& 2\,\chi_0 -\frac{1}{4\,f_\pi^2}
\Big[ \Phi, \Big[ \Phi ,\chi_0 \Big]_+\Big]_+ \,,
\nonumber\\
\chi_0 &=&\frac{1}{3} \left( m_\pi^2+2\,m_K^2 \right)\,1
+\frac{2}{\sqrt{3}}\,\left(m_\pi^2-m_K^2\right) \lambda_8 \;.
\label{chi-sb}
\end{eqnarray}
The parameters $b_D$, $b_F$, and $d_D$ are determined to leading order by the
baryon octet and decouplet mass splitting. The empirical estimates are  
$b_D \simeq (0.064\pm 0.004)$~GeV$^{-1}$, $b_F \simeq -(0.209\pm
0.004)$~GeV$^{-1}$, and $d_D\simeq (0.87\pm 0.04)$~GeV$^{-1}$.  They induce 
well defined meson-octet baryon-octet interaction
vertices. The parameter $b_0$ is related to the pion-nucleon sigma-term
$\sigma_{\pi N} = -2\,m_\pi^2\,(b_D+b_F+2\,b_0)$
with the empirical value \cite{piN-sigterm} $\sigma_{\pi N}=(45\pm 10)$~MeV.

The chiral Lagrangian is a powerful tool once it is combined with appropriate 
power counting rules leading to a systematic approximation strategy. 
In the $\pi N$ sector the $SU(2)$ chiral Lagrangian was successfully applied 
demonstrating good convergence properties of the perturbative 
chiral expansion \cite{Bernard}. In the $SU(3)$ sector the situation is more involved 
due in part to the rather large kaon mass $m_K \simeq m_N/2$. Here the perturbative 
evaluation of the chiral Lagrangian cannot be justified and one must
change the expansion strategy. Rather than expanding directly the scattering amplitude 
one may expand the interaction kernel according to chiral power counting rules 
\cite{LePage}. This is in analogy to the treatment of the $e^+\,e^-$ bound
state problem of QED where a perturbative evaluation of the interaction kernel
can be justified. For details on the formalism how to extract the 
interaction kernel from the relativistic chiral Lagrangian and how to 
then solve for the Bethe-Salpeter scattering equation with a
physical renormalization scheme we refer to \cite{kl-paper}.

\section{Results}

\begin{figure}[t]
\epsfysize=8.5cm
\begin{center}
\mbox{\epsfbox{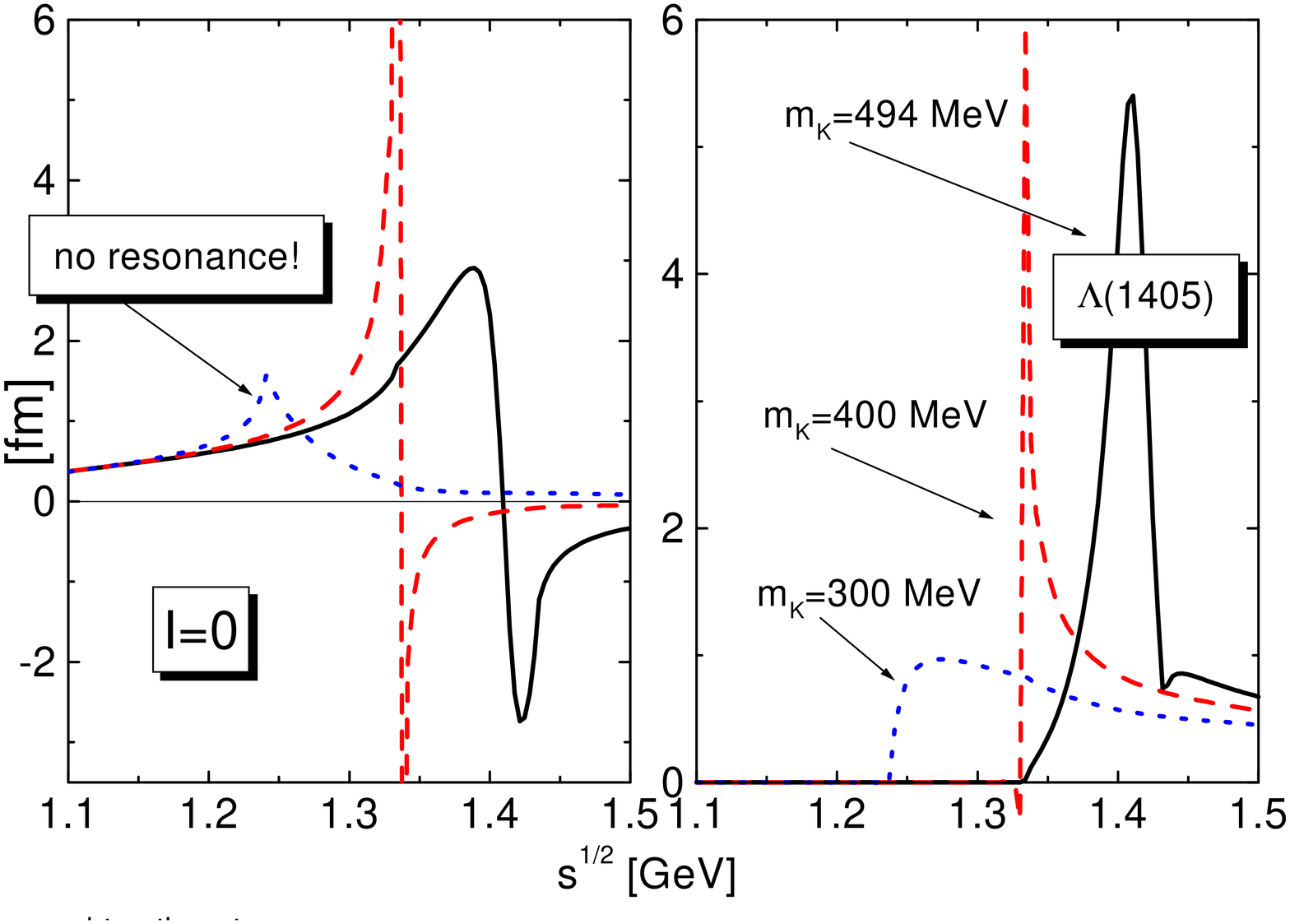}}
\end{center}
\caption{Real (l.h.s.) and imaginary (r.h.s.) part of the isospin zero s-wave $K^-$-nucleon 
scattering amplitude as it follows from the $SU(3)$ Weinberg-Tomozawa interaction term in 
a coupled channel calculation. We use $f_\pi = 93 $ MeV and identify the subtraction point 
with the $\Lambda(1116)$ mass.}
\label{fig1}
\end{figure}

We first discuss the effect of the leading interaction term of the 
chiral $SU(3)$ Lagrangian density suggested long ago by Tomozawa and Weinberg.
If taken as input for the multi-channel Bethe-Salpeter equation,  
properly furnished with a renormalization scheme leading to a subtraction point close to 
the baryon octet mass, a rich structure of the scattering amplitude arises.
In Fig. 1 we show the solution of the multi-channel Bethe-Salpeter as a function of
the kaon mass. For physical kaon masses the isospin zero scattering amplitude shows a 
resonance structure at energies where one would expect the $\Lambda^*(1405)$ resonance.
We point out that the resonance structure disappears as the kaon mass is decreased. 
Already at a hypothetical kaon mass of $300$ MeV the $\Lambda^*(1405)$ resonance is not 
formed anymore. Fig. 1 nicely demonstrates that the chiral $SU(3)$ Lagrangian is
necessarily non-perturbative in the strangeness sector. This confirms the findings of
\cite{Kaiser,Ramos}. However, note that in previous works the $\Lambda^*(1405)$ resonance
is a result of a fine tuned cutoff parameter which gives rise to a different kaon mass
dependence of the scattering amplitude \cite{Ramos}. In our scheme the choice of subtraction 
point close to the baryon octet mass follows necessarily form the compliance of the 
relativistic chiral Lagrangian with chiral counting rules. Moreover, the identification
of the subtraction point with the $\Lambda(1116)$ mass in the isospin zero channel protects the 
hyperon exchange s-channel pole contribution.  

\begin{figure}[t]
\epsfysize=9.0cm
\begin{center}
\mbox{\epsfbox{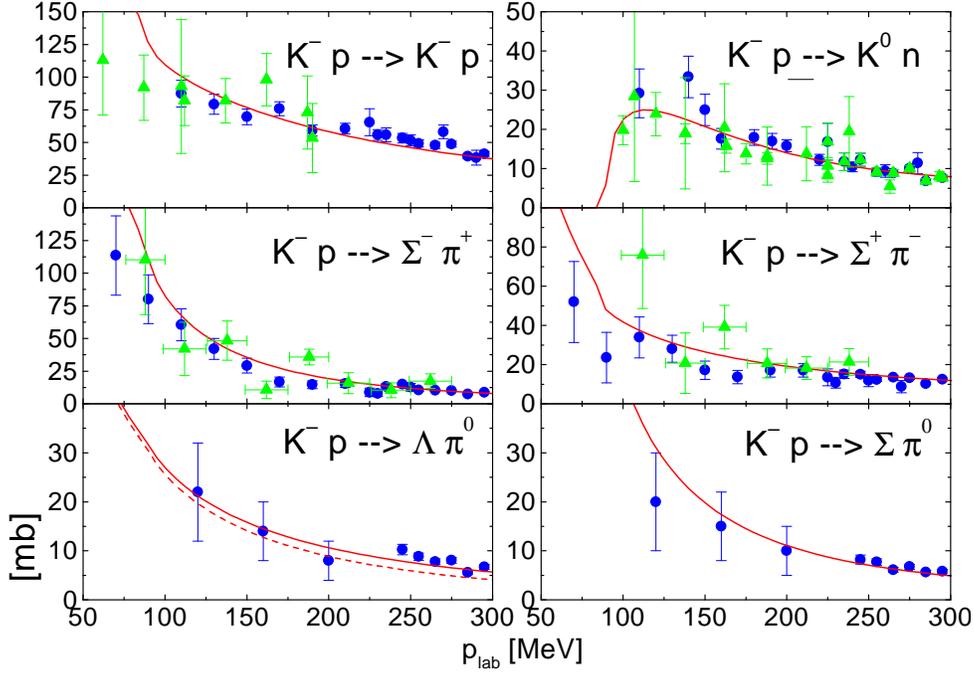}}
\end{center}
\caption{$K^-$-proton elastic and inelastic cross sections. The data are taken from
\cite{Landoldt}. The solid line includes effects of s- and p-waves. The dotted line shows the
s-wave contribution only.}
\label{fig2}
\end{figure}

At subleading order the chiral $SU(3)$ Lagrangian leads to basically 11 free parameters to 
be adjusted to empirical data. Chiral symmetry is found predictive in the $SU(3)$ sector 
since it reduces the number of free parameters. Static $SU(3)$ symmetry alone would 
predict 18 independent terms rather than the 11 chiral terms since 
$8\otimes 8= 1\oplus 8\oplus 8 \oplus 10 \oplus \overline{10}
\oplus 27$. The set of parameters is well determined by elastic 
and inelastic $K^-p$ cross section data together with empirical pion-nucleon threshold 
parameters. In Fig. 2 we present the result of our fit for the elastic and inelastic $K^-p$ 
cross sections. The data set is nicely reproduced including the rather precise data points
for laboratory momenta 200 MeV$<p_{\rm lab}<$ 300 MeV. In Fig. 2 the s-wave contribution to 
the total cross section is shown with a dashed line. Sizeable p-wave contributions are 
found only in the $\Lambda \pi_0$ production cross section. Note that the $\Lambda \pi_0$
channel carries isospin one and therefore provides a valuable constraint on the poorly 
known $K^-$-neutron interaction. 

\begin{figure}[t]
\epsfysize=8.5cm
\begin{center}
\mbox{\epsfbox{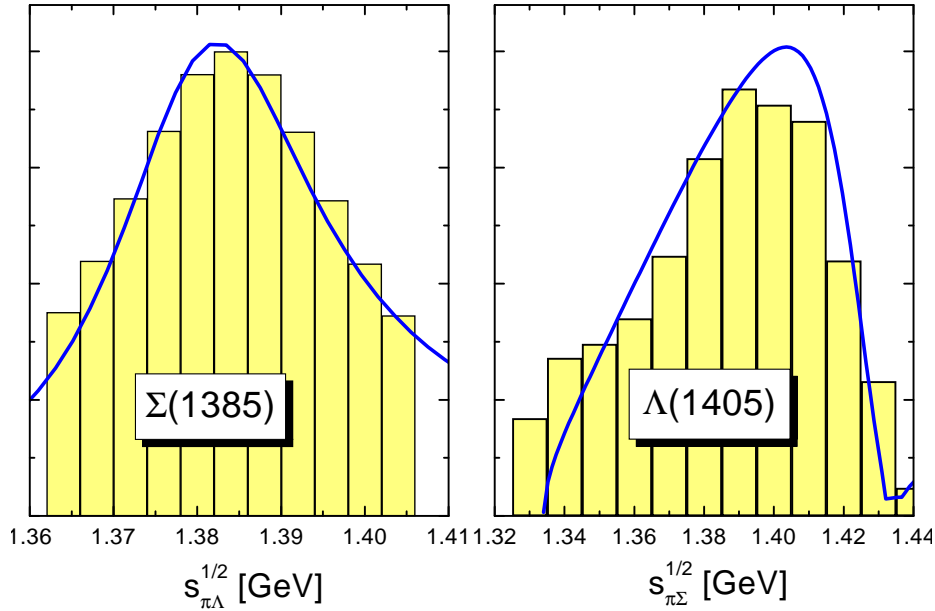}}
\end{center}
\caption{$\Lambda^*(1405)$ and $\Sigma^*(1385)$ resonance mass distributions in arbitrary 
units.}
\label{fig3}
\end{figure}

We worked out two crucial ingredients of a
successful description of the s-wave kaon-nucleon dynamics. First it is found that 
the explicit $\Sigma^*(1385)$ contributions to s- and p-waves are important for a good fit. 
We emphasize that the $\Sigma^*(1385)$ induces s-wave range parameters which 
are $SU(3)$-independent from the chiral range parameters. The situation is different 
from the pion-nucleon system where the isobar induced s-wave range terms can be absorbed
into the chiral range parameters. Second, we find that it is crucial to employ the 
{\it relativistic} chiral Lagrangian. It gives rise to well defined kinematical structures 
in the local 4-point interaction kernel of (\ref{lag-Q}) which leads to a mixing of s-wave 
and p-wave parameters. Only in the heavy-baryon mass limit the parameters decouple into 
the s-wave and p-wave sector. 

\begin{figure}[t]
\epsfysize=9.0cm
\begin{center}
\mbox{\epsfbox{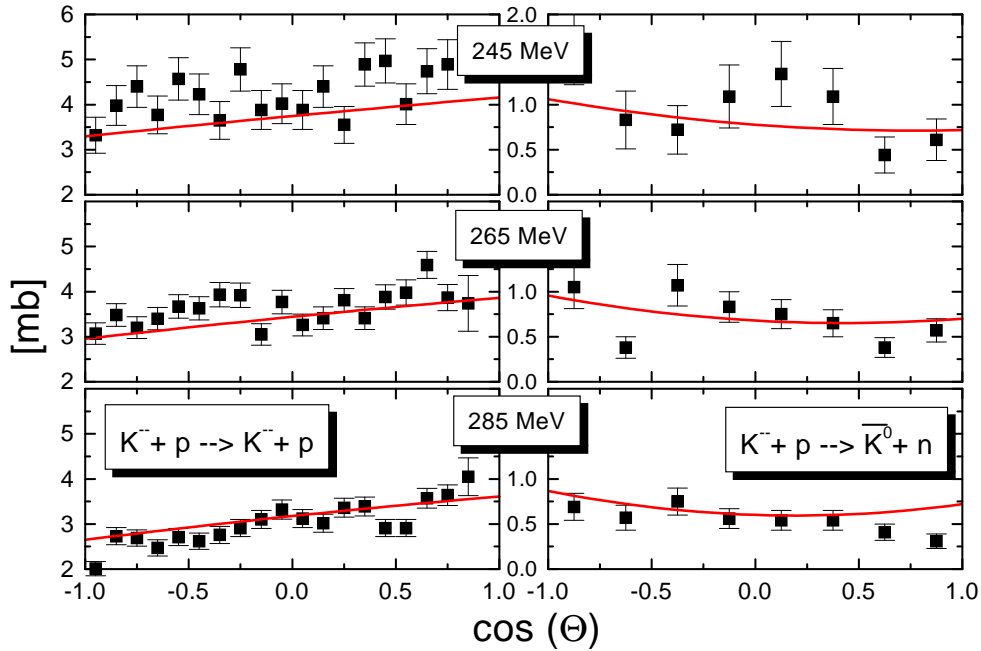}}
\end{center}
\caption{$K^-p$ differential cross sections at $p_{\rm lab}=245$ MeV, 
$p_{\rm lab}=265$ MeV and $p_{\rm lab}=285$ MeV. The data are taken form \cite{mast-ko}.}
\label{fig4}
\end{figure}

In Fig. 3 we show the $\Lambda^*(1405)$ and $\Sigma^*(1385)$ spectral functions measured 
in the reactions $K^-p\rightarrow \Lambda \pi^+\pi^-$ \cite{lb-spec} and 
$K^-p\rightarrow \Sigma^+\pi^-\pi^+\pi^-$ \cite{sig-spec} respectively. We point out 
that even though the spectral form of the $\Lambda^*(1405)$ provides 
an important constraint for the parameter set it does not by itself lead to a stringent 
determination of the isospin zero subthreshold scattering amplitude in the vicinity of the 
$\Lambda^*(1405)$ resonance. This is reflected in the fact that different analyses 
consistent with the low energy data set and the spectral form of the resonance may give 
rise to rather different subthreshold amplitudes \cite{Kaiser,Ramos}.
  
In Fig. 4 we confront our result with available differential cross sections from 
$K^-$ proton scattering. The angular distribution pattern is consistent with a weak 
p-wave contribution. The linear slope in $\cos \theta $ reflects the interference of the 
p-wave contribution with a strong s-wave. The angular distributions of further
inelastic $K^-p$ reactions (not shown) are reproduced with similar quality. We 
also reproduce the empirical $K^-p$ threshold branching ratios \cite{branching-ratios} to 
good accuracy. Here the isospin breaking effects are important.  
In order to unambiguously find a solution we constrain the free parameters to 
also reproduce the empirical s-wave scattering lengths and p-wave scattering volumes
of the pion-nucleon sector. The pion-nucleon threshold parameters 
are evaluated perturbatively to subleading orders in the chiral expansion in accordance 
with effective field theory calculations in the presence of an explicit isobar field. 
Here the small pion mass justifies the perturbative treatment. An accurate 
description of the pion-nucleon threshold parameters is obtained.

\begin{figure}[t]
\epsfysize=9.0cm
\begin{center}
\mbox{\epsfbox{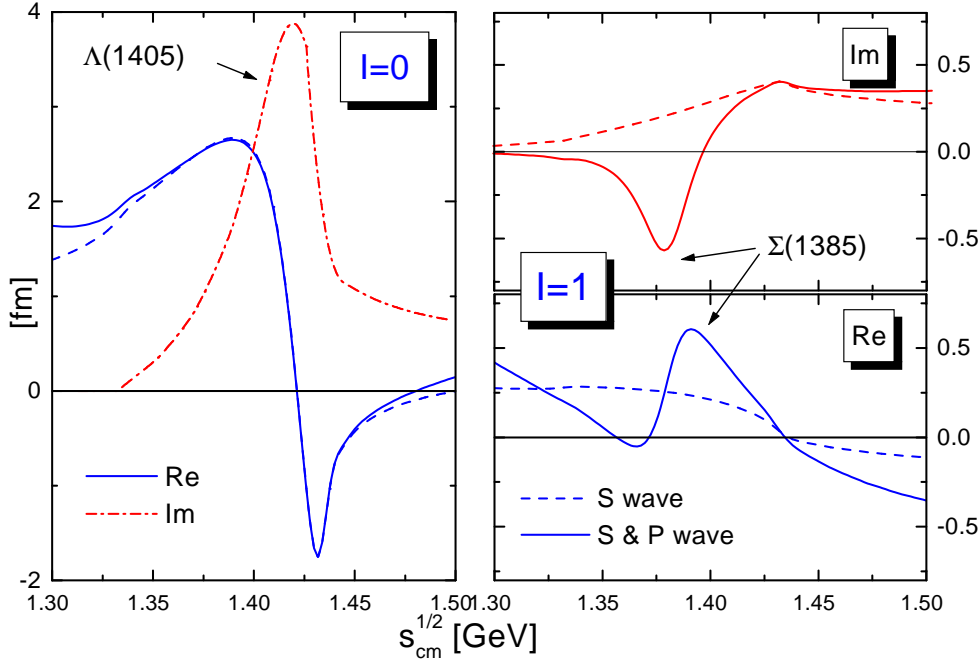}}
\end{center}
\caption{Real and imaginary parts of the
$K^-$-nucleon forward scattering amplitude in the two isospin channels. The full lines
represents the result with s- and p-waves included and  the dashed lines show the s-wave 
contributions to the amplitudes.}
\label{fig5}
\end{figure}

Finally we present our result for the $\bar K$-nucleon forward scattering amplitudes.
In our approach it receives contributions from s- and p-wave channels:
\begin{eqnarray}
f^{(I)}_{KN,\,{\rm forward}}(\sqrt{s}\,)=f^{(I)}_{KN,\,{\rm s-wave}}(\sqrt{s}\,)
+p^2_{KN}\,f^{(I)}_{KN,\,{\rm p-wave }}(\sqrt{s}\,)
\label{}
\end{eqnarray}
where $\sqrt{s}=\sqrt{m_N^2+p_{KN}^2}+\sqrt{m_K^2+p_{KN}^2}$.
As can be seen from Fig. 5 we find sizeable p-wave contributions in the 
subthreshold amplitudes. In particular the $\Sigma(1385)$-resonance dominates the 
isospin one scattering amplitude. Note that p-wave channels contribute with a
positive imaginary part for energies larger than the kaon-nucleon threshold but 
with a negative imaginary part for subthreshold energies. In particular recall that 
a negative imaginary part of a subthreshold amplitude is not forbidden by the optical 
theorem which  relates the imaginary part of the forward scattering amplitudes to 
the total cross section only for energies above threshold. 
We emphasize that our subthreshold amplitudes differ strongly from the ones of 
A.D. Martin \cite{A.D.Martin}. Note that an important and highly model dependent input of  
Martin's dispersion analysis of the $K^-\,p$ forward scattering amplitude was the subthreshold 
amplitude which is not directly constrained by data. 
Clearly, one should investigate constraints from dispersion 
relations in a refined chiral $SU(3)$ analysis.

\section{Discussion and  outlook}

The result for the $K^-$-nucleon scattering amplitudes has interesting consequences 
for kaon propagation in dense nuclear matter. An attractive subthreshold scattering 
amplitude permits a kaon to propagate in nuclear matter with an energy smaller than 
its free space mass.  To leading order in the nuclear density the kaon self energy 
$\Pi_K(\omega, \vec q\,)$ is determined by the s- and p-wave scattering amplitudes 
implicit in Fig. 5. For isospin symmetric nuclear matter 
the low density theorem \cite{dover,njl-lutz} leads to  
\begin{eqnarray}
\Pi_K(\omega ,\vec q\,) &=& -4\,\pi \,\frac{\sqrt{s}}{m_N}\,\Bigg(
f^{({\rm s-wave})}_{KN}(\sqrt{s}\,)
\nonumber\\
&+&\left( \frac{3\,\omega^2}{5\,m_N^2}\,k_F^2
+\vec q\,^2 \right) f^{({\rm p-wave})}_{KN}(\sqrt{s}\,)
\Bigg)\,\rho 
+\cdots
\label{ld-theorem}
\end{eqnarray}
where we imply isospin averaged scattering amplitudes and identify 
$s\simeq (m_N+\omega)^2-\vec q\,^2$. In (\ref{ld-theorem}) we assume 
$\omega ,|\vec q\,|, k_F \ll m_N$ and treat Fermi motion effects in a simplistic 
fashion. The dots in (\ref{ld-theorem}) represent further 
terms of the density expansion. Note that according to (\ref{ld-theorem})
at small momenta $\vec q$ the self energy is {\it not} determined by the forward 
scattering amplitudes  in contrast with the expectation at large momenta. Rather, an 
evaluation of the self energy requires the decomposition of the forward scattering 
amplitudes into its partial wave contributions. 

The spectral function of the kaon is expected to show a complicated structure due to the 
formation of the $\Lambda^*(1405)N^{-1}$ and $\Sigma^*(1385)N^{-1}$ states. This is an 
immediate consequence of the resonance structure seen in the kaon-nucleon scattering 
amplitudes. At small energies $\omega \simeq m_{\Lambda, \Sigma }-m_N$ the spectral 
function may couple also to the hyperon nucleon-hole states $\Lambda N^{-1}$ and 
$\Sigma N^{-1}$ though the spectral weight is expected to 
be rather small \cite{Kolomeitsev}. 

As demonstrated in \cite{sc-lutz} at intermediate nuclear densities $\sim \rho_0$ the 
low density theorem (\ref{ld-theorem}) provides a good 
approximation to the self energy for energies $\omega $ and momenta 
$\vec q$ leading to a $s=(m_N+\omega)^2-\vec q\,^2$ 
sufficiently below or above the resonance structures in the 
scattering amplitudes. On the other hand at zero momentum $\vec 
q=0$ and energies $\omega $ close to the kaon mass, a kinematical 
region of up most importance for the microscopic understanding of 
kaonic atoms, the low density expression is useless already at 
rather small densities $\rho< 0.1\, \rho_0$. This is due to the 
presence of an active small scale determined by the binding energy 
of the $\Lambda^*(1405)$ resonance \cite{sc-lutz}. Since already the vacuum  
scattering amplitude is a rather sensitive function of the kaon mass (see Fig. 1)
one arrives at a self-consistent scheme where the feedback effect of an attractive
kaon spectral function on the in-medium kaon-nucleon scattering process must be taken into 
account \cite{sc-lutz}. Note that though the isospin one amplitude is smaller than 
the isospin zero amplitude its contribution to the kaon self energy in isospin 
symmetric nuclear matter is enhanced by a factor three from  isospin.  
We therefore expect that both the $\Lambda^*(1405)$ and the 
$\Sigma^*(1385)$ resonances need to be included in a self-consistent 
treatment of kaon propagation in nuclear matter. Work in this direction 
is in progress \cite{Korpa}.

Rather than a result of perturbative chiral s-wave dynamics suggested in \cite{Kaplan,CH-Lee} 
we arrive at an intriguing non-perturbative and much richer picture of the chiral nuclear 
kaon dynamics.

\end{document}